\documentclass[fleqn,twoside]{article}
\usepackage[headings]{espcrc2}

\readRCS
$Id: espcrc2.tex,v 1.2 2004/02/24 11:22:11 spepping Exp $
\usepackage{graphicx}
\usepackage[figuresright]{rotating}
\usepackage{amssymb}
\newcommand{\AmS}{{\protect\the\textfont2
  A\kern-.1667em\lower.5ex\hbox{M}\kern-.125emS}}

\title{{\hfill RUB-TPII-08/04}\\ [2cm]
       Pion form factor analysis using NLO analytic perturbation
       theory\thanks{Based on works with A.~P.~Bakulev, S.~V.~Mikhailov,
        K.~Passek-Kumeri\v{c}ki, and W.~Schroers.
               }}

\author{N.~G.~Stefanis\address[RUB]{Institut f\"ur
        Theoretische Physik II, Ruhr-Universit\"at Bochum,
        44780 Bochum, Germany}%
        \thanks{Talk presented at {\it QCD~04}\/, 5-9~July~2004,
        Montpellier, France.}}

\runtitle{Pion form factor analysis using NLO APT}
\runauthor{N.~G.~Stefanis}

\begin{document}

\begin{abstract}
I present results for the pion's electromagnetic form factor in
the spacelike region, which implement the most advanced perturbative
information currently available for this observable in conjunction with
a pion distribution amplitude that agrees with the CLEO data on the
pion-photon transition form factor at the $1\sigma$ level.
I show that using for the running strong coupling and its powers their
analytic versions in the sense of Shirkov and Solovtsov, the obtained
predictions become insensitive to the renormalization scheme and scale
setting adopted.
Joining the hard contribution with the soft part on account of local
duality and respecting the Ward identity at $Q^2=0$, the agreement
with the available experimental data, including expectations from
planned experiments at JLab, is remarkable both in trend and magnitude.
I also comment on Sudakov resummation within the analytic approach.
\vspace{1pc}
\end{abstract}

\maketitle
\section{Introduction}
Driven by the need to understand the pion substructure in terms of
its quark (and gluon) degrees of freedom, an impressive progress has
been achieved in the last few years both from the perturbative side
\cite{MNP98,SSK99,MMP-K02}, as well as from the point of view of a
deeper insight into the underlying nonperturbative dynamics
\cite{BMS01,Kho99}.
In addition, the improved quality of recent \cite{CLEO98} and planned
experiments \cite{JLab01} in conjunction with more sophisticated
data-processing techniques \cite{SY99,BMS02} may soon enable a cleaner
comparison between QCD theory and data.
This talk assesses these issues on the basis of a recent analysis
\cite{BP-KSS04} of the (spacelike) pion's electromagnetic form factor
under the imposition of analyticity of the running strong coupling and
its powers---as far as its factorized part is concerned---including the
soft contribution via local duality.
The discussion focuses on the phenomenological exploitation of these
theoretical concepts, sketching how analytic perturbation theory
(APT)---fixed-order and resummed---can significantly improve the
quality of perturbatively calculable hadronic quantities.
Nonperturbative input enters via the pion distribution amplitude (DA)
\cite{BMS01}, derived from nonlocal QCD sum rules.

\section{Pion's electromagnetic form factor--a role model for APT
         applications}
To start reasoning about ``analytization'' procedures, we have to
make some remarks about the origin of the concept of the analytic
coupling.
Already used in inclusive reactions \cite{DMW96}, the framework of
APT \cite{SS97} has evolved in parallel with other dispersive
techniques \cite{Gru97} to tame the Landau singularity.
A broader framework of analytization in exclusive QCD processes
\cite{SSK99,KS01,BP-KSS04} has permitted to refine---and in some sense
to complexify---progressively the related concepts extending the APT
formalism of \cite{SS97} to hadronic observables with more than one
scheme scales (like $F_{\pi}^{\rm fact}(Q^2)$ at NLO), and including
ERBL\ evolution \cite{ERBL79}, and also Sudakov resummation pertaining
to both logarithms and power corrections.
The two main objectives of this framework are (i) the augmentation of
exclusive amplitudes with IR protection against the Landau ghost and
(ii) the improvement of perturbative QCD calculations by rendering
them insensitive to the choice of the renormalization scheme and scale
adopted.
This allows to obtain predictions with significantly reduced
theoretical bias, permitting this way a much cleaner comparison with
experimental data.

The factorized spacelike pion's electromagnetic form factor in NLO
analytic perturbative QCD reads
\begin{equation}
F_{\pi}^{\rm fact}(Q^2; \mu_{\rm R}^{2})
  =  F_{\pi}^{\rm LO}(Q^2;   \mu_{\rm R}^{2})
   + F_{\pi}^{\rm NLO}(Q^2;  \mu_{\rm R}^{2})
\label{eq:totfactff}
\end{equation}
%Eq (1) Total factorized FF up to NLO
with the LO and NLO terms given by
\begin{eqnarray}
  \!\!\!\!\!\!\!F_{\pi}^{\rm LO}(Q^2;\mu_{\rm R}^{2})\!\!\!\!
&\!\!\!\!\!\!\!\!=\!\!\!\!\!\!\!\!\!&
  \alpha_{\rm s}(\mu_{\rm R}^{2})\, {\cal F}_{\pi}^{\rm LO}(Q^2) \, ,\\
  Q^2 {\cal F}_{\pi}^{\rm LO}(Q^2)
&\equiv&
  8\,\pi\,f_{\pi}^2\,
    \left[1 + a_2^{\rm D,NLO}(Q^2)\right.\\ \nonumber
&+& \!\!\!\!\left. a_4^{\rm D,NLO}(Q^2)\right]^2 \, ;
%\label{eq:Q2pffLO}
\end{eqnarray}
%%Eq (2) LO pion FF in terms of calligraphic FF
%%Eq (3) LO calligraphic pion FF with NLO evolution of pion DA
\begin{eqnarray}
  F_{\pi}^{\rm NLO}(Q^2;\mu_{\rm R}^{2})\!\!\!\!
&\!\!\!\!\!=\!\!\!\!\!&\!\!\!\!
    \frac{\alpha_{\rm s}^2(\mu_{\rm R}^{2})}{\pi}\,
    \left[{\cal F}_\pi^{\rm D,NLO}(Q^2;\mu_{\rm R}^{2}) \right.
\\ \nonumber
&+& \!\!\!\!\left. {\cal F}_\pi^{\rm ND,NLO}(Q^2;N_{\rm Max}=\infty)
    \right] \, ,
\label{eq:Q2pffNLO}
\end{eqnarray}
%%Eq (4) NLO pion FF in terms of calligraphic pion FF
where $N_{\rm Max}$ marks the maximal number of Gegenbauer harmonics
taken into account and the quantities in calligraphic notation are
provided in \cite{BP-KSS04}.
Note that these expressions take into account the NLO evolution of
the pion distribution amplitude and hence contain diagonal (D) as well
as (the NLO term) non-diagonal (ND) components.
The effects of the LO evolution are crucial \cite{MNP98}, while those
of the NLO are relatively of less importance.
This allows us to set \cite{BP-KSS04}:
$
  a_n^{\rm D,NLO} \to a_n^{\rm D,LO}
 \;\, \mbox{and} \;\,
  a_n^{\rm ND,NLO} \to  0.
$
For a detailed exposition of this material, see \cite{BP-KSS04}.

%%%%%%%%%%%%%%%%%%%%%%%%%%% F I G U R E  1 %%%%%%%%%%%%%%%%%%%%%%%%%%%%
\begin{figure}[t]%[!thb]
\centerline{\includegraphics[width=1.0%
\columnwidth]{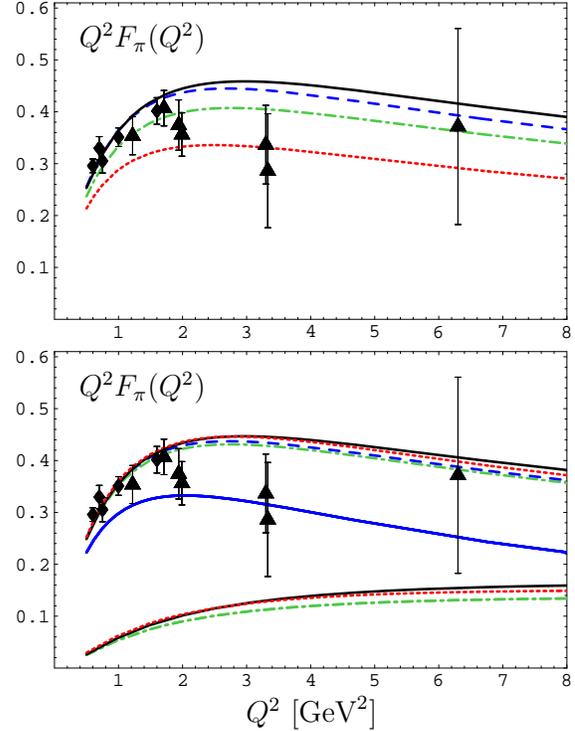}}
\vspace*{-10mm}
\caption[*]{\footnotesize Predictions for $Q^2F_\pi(Q^2)$ using APT and
  the BMS\ pion DA \protect\cite{BMS01} in conjunction with the ``Naive
  Analytic'' (top) and ``Maximally Analytic'' (bottom) procedures
  \protect\cite{BP-KSS04}:
  $\overline{\strut{\rm MS}}$ scheme and $\mu_{\rm R}^{2}=Q^2$
  (dashed line); BLM\ (dotted line);
  $\overline{\strut{\rm BLM}}$ (solid line); $\alpha_V$-scheme
  (dash-dotted line).
  The single solid line in the lower panel shows the prediction for
  the soft form-factor part; below this, the corresponding hard
  contributions are also displayed.
\label{fig:ff-naiv-max}}
\vspace*{-7mm}
\end{figure}
%%%%%%%%%%%%%%%%%%%%%%%%%%%%%%%%%%%%%%%%%%%%%%%%%%%%%%%%%%%%%%%%%%%%%%%

The analytization of $F_{\pi}^{\rm fact}(Q^2)$ means to replace the
running coupling and its powers by analytic expressions.
We consider here two different analytization procedures in parity:\\
(i) \emph{Naive Analytization} \cite{SSK99,BP-KSS04} replaces in
$F_{\pi}^{\rm fact}$ the strong coupling and its powers by the
analytic coupling $\bar{\alpha}_{\rm s}$ \cite{SS97} and its powers
$\left[\bar{\alpha}_{\rm s}^{(N)}(\mu_{\rm R}^{2})\right]^n$, i.e.,
\begin{equation}
  \mathbb{A}_{\rm naive}[\alpha_{\rm s}^{(N)n}(Q^2)]
=
  \bar{\alpha}_{\rm s}^{(N)n}(Q^2) \, ,
\label{eq:naiv}
\end{equation}
%%Eq (5) Naive Analytization of coupling and its powers
amounting at NLO to
\begin{eqnarray}
&& \!\!\!\!\!\!\!\!\!\!\!\!
\left[F_{\pi}^{\rm fact}(Q^2; \mu_{\rm R}^{2})\right]_{\rm NaivAn}
=\bar{\alpha}_{\rm s}^{(2)}(\mu_{\rm R}^{2})\,
    {\cal F}_{\pi}^{\rm LO}(Q^2) \\ \nonumber
&&+ \frac{1}{\pi}\,
      \left[\bar{\alpha}_{\rm s}^{(2)}(\mu_{\rm R}^{2})\right]^2\,
      {\cal F}_{\pi}^{\rm NLO}(Q^2;\mu_{\rm R}^{2})\,.
\label{eq:pffNaivAn}
\end{eqnarray}
%%Eq (6) Imposition of Naive analytization on NLO pion FF
(ii) \emph{Maximal Analytization} \cite{BP-KSS04} associates to
the powers of the running coupling their own dispersive images,
trading this way the usual power series expansion for a non-power
functional expansion \cite{SS97} to get
\begin{equation}
  \mathbb{A}_{\rm max}[\alpha_{\rm s}^{(N)n}(Q^2)]
=
  {\cal A}_{n}^{(N)}(Q^2) \, ,
\label{eq:max}
\end{equation}
%%Eq (7) Maximal Analytization of coupling and its powers
where $N$ is the number of loops and $n$ the index of expansion.
This entails at the two-loop level
\begin{eqnarray}
&& \!\!\!\!\!\!\!\!\!\!\!\!
\left[F_{\pi}^{\rm fact}(Q^2; \mu_{\rm R}^{2})\right]_{\rm MaxAn}
 \ =\bar{\alpha}_{\rm s}^{(2)}(\mu_{\rm R}^{2})\,
    {\cal F}_{\pi}^{\rm LO}(Q^2) \\ \nonumber
&&+ \frac{1}{\pi}\,
      {\cal A}_{2}^{(2)}(\mu_{\rm R}^{2})\,
      {\cal F}_{\pi}^{\rm NLO}(Q^2;\mu_{\rm R}^{2})\, ,
\label{eq:pffMaxAn}
\end{eqnarray}
%%Eq (8) Imposition of Maximal Analytization on NLO pion FF
with $\bar{\alpha}_{\rm s}^{(2)}$ and
${\cal A}_{2}^{(2)}(\mu_{\rm R}^{2})$
being the 2-loop analytic images of $\alpha_{\rm s}^{(2)}(Q^2)$
and $\left(\alpha_{\rm s}^{(2)}(Q^2)\right)^2$,
respectively.
Studying $F_{\pi}^{\rm fact}(Q^2)$ beyond the LO requires an optimal
renormalization scheme and scale setting in order to minimize the
influence of higher-order loop corrections and avoid dependence of the
results on the particular renormalization scheme and scales adopted.
An in-depth analysis of these issues was carried out in
\cite{BP-KSS04}, where estimates of $F_{\pi}^{\rm fact}(Q^2)$ in
conventional QCD perturbation theory within the
$\overline{\strut{\rm MS}}$ scheme with various scale settings were
contrasted to analogous results obtained in APT.
To confront these predictions with experimental data
(Fig.\ \ref{fig:ff-naiv-max}), the soft nonfactorizable contribution
was incorporated via local duality (LD) and the Ward identity at
$Q^2=0$ was implemented by a power-behaved pre-factor in order to
ensure that each of these two contributions was evaluated in its own
momentum region of validity.
Hence, we have
\begin{eqnarray}
  F_{\pi}(Q^{2};\mu_{\rm R}^{2})
  =  F_{\pi}^{\rm LD}(Q^{2})
  +  \hat{F}_{\pi}^{\rm fact}(Q^2;\mu_{\rm R}^{2})
\label{eq:Q2Pff}
\end{eqnarray}
%%Eq (9) Total pion form factor
with
\begin{eqnarray}
 \hat{F}_{\pi}^{\rm fact}(Q^2;\mu_{\rm R}^{2})
= \left[\frac{Q^2}{2s_0^{\rm (2)}+Q^2}\right]^2\!
       F_{\pi}^{\rm fact}(Q^2;\mu_{\rm R}^{2})
\label{eq:Fpi-Mod}
\end{eqnarray}
%%Eq (10) Interpolating expression from low to high Q^2 for
%%       factorizable form factor
and $s_{0}^{(2)}\simeq 0.6$~GeV${}^2$ (for more details, see
\cite{BP-KSS04}).
%%%%%%%%%%%%%%%%%%%%%%%%%%% F I G U R E  2 %%%%%%%%%%%%%%%%%%%%%%%%%%%%
%\vspace*{-1mm}
\begin{figure}[!thb]
\centerline{\includegraphics[width=1.0%
\columnwidth]{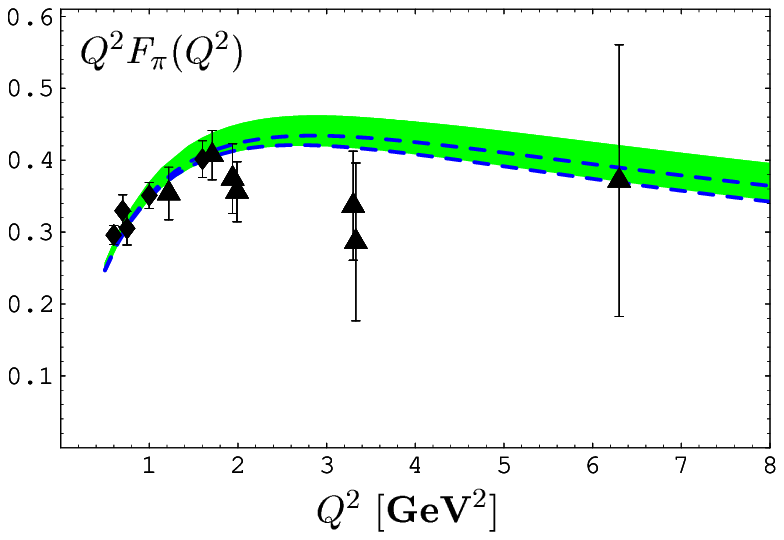}}
\vspace*{-5mm}
\caption[*]{\footnotesize Predictions for $Q^2F_{\pi}(Q^2)$ calculated
   with the ``Maximally Analytic '' procedure and the pion DAs derived
   with nonlocal QCD sum rules \protect\cite{BMS01} (shaded strip). The
   broken lines denote the region accessible to the asymptotic pion DA.
   The experimental data are taken from \protect{\cite{JLab01}}
   (diamonds) and \protect\cite{FFPI73} (triangles).
\label{fig:ff-strip}}
\vspace*{-7mm}
\end{figure}
%%%%%%%%%%%%%%%%%%%%%%%%%%%%%%%%%%%%%%%%%%%%%%%%%%%%%%%%%%%%%%%%%%%%%%%
The main phenomenological upshot of the presented analysis is shown in
Fig.\ \ref{fig:ff-strip}.
It is interesting to observe from this figure that the theoretical
error bounds (shaded strip) induced by the nonperturbative
determination of the pion DA within the nonlocal QCD sum-rules picture
are much smaller than those of the current high-momentum experimental
data.
This situation may, however, dramatically improve in a couple of years
when the approved upgrade of CEBAF@JLab to an energy of $12$~GeV will
start delivering high-precision data up to momentum transfers of about
$6$~GeV$^2$ (see second entry in \cite{JLab01}).
Another striking observation from this figure is that the form-factor
predictions obtained with the double-humped BMS\ pion DA are only
slightly larger than those following from the single-peaked asymptotic
DA (area within the two broken lines).
Hence, it becomes evident that what counts for the form factor is not
the central region of longitudinal momenta around $x\simeq 1/2$ of the
pion DA, but its endpoint regions $x\sim 0,1$
\cite{BMS01,BMS02,BP-KSS04}, as pointed out before in \cite{SSK99}.
If the endpoint region is suppressed, then the form-factor prediction
does not get artificial enhancement that may jeopardize the
perturbative treatment.
This suppression is controlled by the nonlocality of the scalar quark
condensate, parameterized by the average quark virtuality
$\lambda_q^2$ in the vacuum, with theoretical estimates in the
range $(0.4 - 0.5)~\mbox{GeV}^2$ \cite{BM02} and a preferable
value of $0.4$~GeV$^2$ extracted in \cite{BMS02} from the CLEO
data.
%%%%%%%%%%%%%%%%%%%%%%%%%%% F I G U R E  3 %%%%%%%%%%%%%%%%%%%%%%%%%%%%
\begin{figure}[!thb]
%\vspace*{-10mm}
 \centerline{\includegraphics[width=1.0\columnwidth]{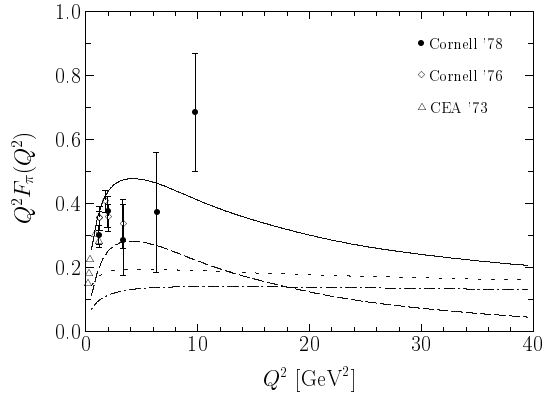}}
  \caption[*]{\footnotesize Predictions for the spacelike pion form
  factor including Sudakov effects within the Naive Analytization
  framework \protect\cite{SSK99}. LO calculation---dash-dotted line;
  NLO result---dotted line; soft-overlap contribution---dashed line.
  The solid line represents the sum of the NLO hard contribution and
  the soft one.
\label{fig:ffpi-sud}}
%\vspace*{-7mm}
\end{figure}
%%%%%%%%%%%%%%%%%%%%%%%%%%%%%%%%%%%%%%%%%%%%%%%%%%%%%%%%%%%%%%%%%%%%%%%

Another means to suppress the endpoint region is provided by the
Sudakov form factor \cite{LS92} to which we now turn our attention.
Sudakov resummation in conjunction with ``Naive Analytization'' was
first considered in \cite{SSK99} in connection with the asymptotic
pion DA and Fig.\ \ref{fig:ffpi-sud} shows the prediction for the
form factor based on a NLO calculation similar to that in
\cite{BP-KSS04}.

\section{Conclusions}
I have discussed a cutting-edge analysis \cite{BP-KSS04} of the
electromagnetic pion form factor, representing a confluence of
advantages in QCD, ranging from non-power series fixed-order
\cite{SS97} and resummed analytic perturbation theory \cite{SSK99}
to an improved CLEO data processing \cite{SY99,BMS02} and to
nonperturbative modelling of the pion distribution amplitude via
nonlocal condensates \cite{BMS01}.
It appears that the principle of Maximal Analytization \cite{BP-KSS04}
of hadronic observables in conjunction with QCD perturbation theory
helps to offset the renormalization-scheme and scale-setting
dependence---unavoidable in the conventional power-series perturbative
expansion---already at NLO.
The crucial observation on the nonperturbative side \cite{BMS02} is
that the CLEO \cite{CLEO98} and the CELLO \cite{CELLO91} data on the
pion-photon transition and also the JLab\ data \cite{JLab01} on the
pion's electromagnetic form factor can be best described by the doubly
peaked but endpoint-suppressed BMS\ pion distribution amplitude
\cite{BMS01} with all other known model distribution amplitudes being
relatively disfavored by one or the other set of experimental data at
least at the $2\sigma$ level \cite{BP-KSS04,BMS04}.

I am confident that the analytic perturbative approach presented
here---to be seen in conjunction with several other applications
discussed elsewhere \cite{SS97}---is a key step towards achieving a
better control over perturbative expansions, in particular, in the low
$Q^2$ domain---especially after integrating more accurately resummation
techniques under (Maximal) Analytization.

\section*{Acknowledgments}
I would like to thank J.~Papavassiliou and D.~V.~Shirkov for useful
discussions and the Deutsche Forschungsgemeinschaft for a travel grant.

\end{document}